\documentclass{article}
\usepackage{amsmath,graphicx}
\newtheorem{theorem}{Theorem}[section]   
   
\newtheorem{remark}{Remark}[section]

\usepackage{amssymb,amsmath,mathrsfs,mathtools}
\usepackage[dvips]{color}
\usepackage{enumerate}
\usepackage[utf8]{inputenc}
\usepackage[english]{babel}
\usepackage{graphics,graphicx}
\usepackage{latexsym}
\usepackage{times}

\newcommand{\n}{\frac{1}{n}\sum_{i=1}^n}
\newcommand{\argmin}{\mathop{\mathrm{arg\,min}}}
\newcommand{\tr}{\mathrm{tr}}
\newcommand{\tran}{\mathrm{T}}
\newcommand{\R}{\mathbb{R}}
\newcommand{\Vhh}{\widehat{V}_1(Z)}
\begin{document}

\title{Breakdown points of penalized and  hybrid \\ M-estimators of covariance}
\author{\begin{tabular}{ccc}  David E. Tyler  & \mbox{ }  & Mengxi Yi \\
\normalsize{Department of Statistics} &  & \normalsize{School of Statistics} \\
\normalsize{Rutgers, The State University of New Jersey} &  & \normalsize{University of International Business and Economics} \\
\normalsize{Piscataway, NJ, U.S.A.} &  & \normalsize{Beijing, China} \\
\normalsize{dtyler@stat.rutgers.edu} &  & \normalsize{myi@uibe.edu.cn}
\end{tabular}
}

\maketitle

\abstract{
We introduce a class of  hybrid M-estimators of multivariate scatter which, analogous to the popular spatial sign covariance matrix (SSCM), 
possess high breakdown points. We also show that the SSCM can be viewed as an extreme member of this class.
Unlike the SSCM, but like the regular M-estimators of scatter, this new class of estimators takes into account the shape of the contours of the data cloud for downweighting observations. \\[4pt]
}

\section{Introduction}  

Though being the most efficient at normal distribution, the sample covariance matrix is well known to be non-robust in the sense that it is highly influenced 
by outliers and has relatively low efficiency at distributions which have longer tails than the multivariate normal distribution. A variety of 
of robust alternatives to the sample covariance matrix have been proposed, with the multivariate M-estimators \cite{Maronna:1976, Huber:1977} being 
one of the earlier ones introduced.  A drawback to the multivariate M-estimators, though, is they have relatively low breakdown points in
higher dimensions, specifically no greater than $1/q$, with $q$ being the dimension of the data. Consequently, high-breakdown point estimators of scatter 
were developed. These include the minimum volume ellipsoid (MVE) and minimum covariance determinant (MCD) estimators \cite{Rousseeuw:1985},  
S-estimators \cite{Davies:1987, Lopuhaa:1989}, projection-based scatter estimators \cite{Donoho-Gasko:1992, Maronna:1992}, and 
MM-estimators \cite{Tatsuoka-Tyler:2000}. 

All the above estimators are affine equivariant. For a $q$-dimensional sample $X=\{x_1,\ldots, x_n\}$, a multivariate scatter 
statistic, say $V(X)$, is said to be \emph{affine equivariant} if  an affine transformation on the data $x_i \to Ax_i +a$, $i = 1, \ldots, n$, induces 
the  transformation on the scatter statistic  $V(A) \to A V(X) A^{\tran}$, for $a\in\R^q$ and $A\in GL(q)$, the group of 
all $q\times q$ nonsingular matrices. 

As shown in \cite{Davies:1987}, the finite sample contamination breakdown point of any affine equivariant scatter statistics is at most $(n-q+1)/(2n-q+1)$ 
for any data set in general position, which is relatively low when $n$ is not much larger than $q$. Furthermore, unlike the M-estimators, affine equivariant 
high breakdown point estimators tend to be computationally intensive, especially for large $q$. They are typically computed via approximate or 
probabilistic algorithms. To handle these shortcomings, scatter estimators that sacrifice affine equivariance have been proposed, such as the spatial sign 
covariance matrix (SSCM) \cite{Locantore:1999, Visuri:2000}. The SSCM is an appealing candidate because of its simplicity. Not only does it provide 
a robust scatter estimator with a high breakdown point,  but  it is also computationally feasible for large $q$.

The lack of affine equivariance of SSCM can result in a severe loss of statistical power when the true distribution deviates strongly from sphericity; see 
\cite{Croux:2002, Magyar:2014}. In this paper, we show that penalized M-estimators of scatter, though, can also have high breakdown points. In
addition, we introduce a standardized version of the penalized M-estimators, which we refer to as hybrid M-estimators, and show that the SSCM can 
be viewed as an extreme element of this class, namely as the tuning parameter goes to its upper limit. The other extreme element of the class, 
i.e.\ when the tuning parameter is zero, corresponds to a regular M-estimator of multivariate scatter. Unlike the SSCM, the hybrid M-estimators can also be 
tuned somewhat for efficiency while maintaining a high breakdown point. 

The rest of this paper is organized as follows. We begin by reviewing the notion of the breakdown point and the definition and properties of the spatial sign 
covariance matrix in sections \ref{Sec:bd} and \ref{Sec:SSCM} respectively.  The main results of the paper are given in section \ref{Sec:robust}, which 
gives results on the breakdown points of penalized M-estimators of scatter, and section \ref{Sec:hybrid}, which introduces the hybrid M-estimators of 
scatter. Some concluding remarks are made in section \ref{Sec:conclude}.

\section{Preliminary concepts} \label{Sec:Prelim}

\subsection{Breakdown due to contamination} \label{Sec:bd}
The breakdown point of an estimator represents a measure of its \emph{global} robustness. Since its formal introduction in 
 \cite{Hampel:1971}, a variety of definitions for the breakdown have been proposed. Here we use the concept of the finite sample contamination
 breakdown point,  defined in \cite{Donoho:1982,Donoho-Huber:1983}. Roughly translated, this corresponds to the proportion of bad data in a sample
 that can be tolerated by an estimator before making the estimator arbitrarily bad. The formal definition is as follows.

Suppose the data set $X=\{x_1,\cdots,  x_n\}$ is contaminated by adding $m$ arbitrary data points $Y=\{y_1,\cdots, y_m\}$. This 
produces an $\epsilon$-contaminated sample $Z = X \cup Y$ consisting of a fraction $\epsilon_m=m/(n+m)$ of arbitrary bad values. Let
$T(\cdot)$ represent a statistic, i.e.\ a function of the data. Next, define an unbounded \emph{bias} function $B(T(X),T(Z))$, which measures
the discrepancy between  $T(X)$ and $T(Z)$ due to the contamination $Y$.  For example, for a location statistic, a natural choice for the 
bias function is $B(t_1,t_2) = \|t_1 - t_2\|$. For $\epsilon=\epsilon_m$, the maximum bias at $X$ caused by $\epsilon$-contamination is 
given by 
\[
b(\epsilon;X)=\begin{cases}
\sup_{Y} B\{T(X),T(Z)\} ,\quad & Z  \in \mathcal{S}_m(X)\\
\infty,\quad & Z \notin \mathcal{S}_m(X)
\end{cases},
\]
where $\mathcal{S}_m(X)=\{Z \,|\, T(Z) \text{ exists}\}$. Breakdown is said to occur under $\epsilon$-contamination if $b(\epsilon;X)=\infty$.
The finite sample contamination breakdown point of $T(X)$ at $X$ is then defined to be
$\epsilon^*(X;T)=\min_{1 \le m \le n}\left\{m/(n+m)\,|\,b(\epsilon; X)=\infty\right\}.$

For a $q$-dimensional scatter statistic $V(\cdot)$, i.e. one whose range is the set of positive definite matrices, a natural choice for the bias 
function is the Riemannian metric $B(V_1,V_2) = ||\log(V_1^{-1/2}V_2V_1^{-1/2})||_F$. In this case, breakdown implies either the  
largest eigenvalue of $V(Z)$ can be arbitrarily large, the smallest eigenvalue of $V(Z)$ can be arbitrarily close to zero, or $V(Z)$ does
not exist for some $Y$.

\subsection{Spatial sign covariance matrix} \label{Sec:SSCM}
In this section, we briefly review some properties of spatial sign covariance matrix. For more details, we refer the reader to \cite{Durre:2014},
\cite{Magyar:2014} and \cite{Visuri:2000}.  For $x\in\R^q$, the spatial sign function is defined as $S:\R^q\to\R^q$ with $S(x)=x/\|x\|$
for $x\ne 0$ and $S(0)=0$, where $\|\cdot\|$ denotes the Euclidean norm in $\R^q$. The spatial sign corresponds to the unit vector in the 
direction of $x$, and it is equal to the univariate sign function when $q=1$. The sample spatial sign covariance matrix is then defined as
\[\mathcal{S}_n(X)=\n S(x_i-\mu_n)S(x_i-\mu_n)^{\tran},\]
where $\mu_n=\argmin_{\mu\in\R^q}\sum_{i=1}^n\|x_i-\mu\|$ is the spatial sign median. Note that $\n S(x_i-\mu_n)=0$, and so
$\mathcal{S}_n(\cdot)$ corresponds to the sample covariance matrix computed from the spatial signs of the data. The SSCM is scale and location 
invariant and also orthogonally equivariant, i.e. the data transformation $x_i \to \alpha Q x_i +a$, $i = 1, \ldots, n$  induces  the transformation  
$\mathcal{S}_n(X) \to  Q \mathcal{S}_n(X) Q^{\tran}$, for $\alpha \in \R$, $a\in\R^q$  and $Q\in O(q)$, the group of orthogonal matrices of order $q$.
The breakdown point of the SSCM is approximately $1/2$. However, if $\mu_n$ is replaced with a fixed location $\mu$, then the breakdown
point becomes one, i.e.\ it cannot break down.

When the data $X$ represents a random sample from $x$ having distribution $F$, then under mild conditions on $F$ the sample SSCM is strongly 
consistent for its functional or population value
$\mathcal{S}(F)=E_F\{S(x-\mu)S(x-\mu)^{\tran}\}$,
where $E_F$ refers to expectation under the distribution $F$,  and $\mu=\mu(F)=\argmin_{\mu\in\R^q}E_F(\|x-\mu\|-\|x\|)$ is the population 
spatial median. When $F$ is an elliptical distribution with finite
second moments, then the SSCM is known to have the same eigenvectors as those of the population covariance matrix, with the corresponding 
eigenvalues having the same order and multiplicities.  However, as shown in \cite{Durre:2016}, the eigenvalues of the population SSCM tend to be
less separated than the eigenvalues of the population covariance matrix. Hence the SSCM can be viewed as shrinking the covariance matrix 
towards proportionality to the identity matrix.

This last property suggests that the SSCM may be related to penalized estimators of the covariance matrix.  We review in the next section the
penalized M-estimators of multivariate scatter, as well as present new results on their breakdown point, with their relationship to 
the SSCM discussed in section \ref{Sec:hybrid}.

\section{Breakdown properties of the penalized M-estimators} \label{Sec:robust}

When one has insufficient data, in that the sample size is small relative to the dimension of the data,  regularization or penalization methods are natural to 
consider. In particular, a ridge approach for regularizing the sample covariance matrices is considered in \cite{Warton:2008}, whereas the existence, 
uniqueness and computation of a regularized Tyler's estimator is studied in \cite{Pascal-etal:2014,Sun:2014}. Penalized versions of the M-estimators 
of multivariate scatter where proposed in  \cite{Ollila-Tyler:2014}, with special attention given to penalizing the trace of the precision matrix. For
general penalties, the existence, uniqueness and computation of the penalized M-estimators of scatter are treated in \cite{Duembgen-Tyler:2016}.  There 
appears to be little literature, though, regarding the robustness of a penalized M-estimator of multivariate scatter.

Assume that the q-dimensional data $x_1, \ldots, x_n$ has been robustly centered, e.g.\ by marginal medians or the spatial median, or  
by a known center. Alternatively, the data may be replaced with their pairwise differences. A penalized M-estimator of scatter is then defined as the
matrix $\widehat{\Sigma}$ which minimizes, 
\begin{equation}  \label{eq:M-obj}
L_\rho(\Sigma; \eta) = \frac{1}{n} \sum_{i=1}^n \rho(x_i^{\tran} \Sigma^{-1} x_i) + \log\{ \det(\Sigma)\} + \eta \Pi(\Sigma), \end{equation}
over all $\Sigma > 0$, the set of positive definite symmetric matrices of order $q$. Here $\rho: \R^+ \to \R$, $\Pi(\Sigma)$ denotes a non-negative 
penalty function and $\eta \ge 0$ is the penalty tuning parameter. For $\eta = 0$, \eqref{eq:M-obj} reduces to the objective or loss function associated 
with an M-estimator of scatter based on a given $\rho$-function. Hereafter, we assume $\rho(s)$ is continuously 
differentiable, and let $u(s)=\rho'(s)$, with the ``influence function" $\psi(s)=su(s)$ being non-decreasing.  

We first consider the trace precision matrix penalty $\Pi(\Sigma)=\tr(\Sigma^{-1})$, which has the effect of heavily penalizing $\Sigma$ when it is 
nearly singular. Properties of this penalty has been studied in \cite{Ollila-Tyler:2014, Warton:2008}. In particular, it is shown in \cite{Ollila-Tyler:2014}
that, for any $\eta \ge 0$, any critical point $\Sigma>0$ of the penalized loss function \eqref{eq:M-obj} is a solution to the penalized M-estimating 
equation
\begin{equation}\label{eq:OT}
\Sigma=\frac{1}{n} \sum_{i=1}^n u(x_i^{\tran} \Sigma^{-1} x_i)x_ix_i^{\tran}+\eta  I_q.
\end{equation}
Furthermore, if $\rho(s)$ is bounded below, then \eqref{eq:OT} has a unique solution for $\eta > 0$ regardless of the data set. When $\eta = 0$, i.e.\ an
unpenalized M-estimator of scatter, it is known some conditions on the data set are needed to guarantee exisitence \cite{Kent-Tyler:1991}. 

Similar results hold when using the Kullback-Leibler penalty function $\Pi_{KL}(\Sigma)=\tr(\Sigma^{-1})+\log\det(\Sigma)$. 
Namely, the corresponding M-estimating equation is given by
\begin{equation} \label{eq:KL}
\Sigma=(1-\gamma) \frac{1}{n} \sum_{i=1}^n u(x_i^{\tran} \Sigma^{-1} x_i)x_ix_i^{\tran}+ \gamma  I_q, 
\end{equation}
where $\gamma = \eta/(1+\eta)$. Again,  if $\rho(s)$ is bounded below, then \eqref{eq:KL} has a unique solution for $0 <\gamma \le 1 $ 
regardless of the data set.

If  $u(s) = \kappa/s$ then, as with Tyler's M-estimator \cite{Tyler:1987}, the resulting penalized M-estimator is distribution-free over the class of elliptical distributions. This follows since  the distribution of $x/\|x\|$ has a fixed distribution, namely the angular Gaussian distribution with parameter
$\Sigma$, whenever $x$ has any elliptical distribution centered at $0$ and with shape matrix $\Sigma$. However, the corresponding $\rho$-function, given by  $\rho(s) = \kappa \log(s)$, is not bounded below. For this case, though, it is also shown in \cite{Ollila-Tyler:2014} that a unique 
solution to \eqref{eq:OT} exists regardless of the data whenever $\kappa < 1$. By relating equations \eqref{eq:KL} to \eqref{eq:OT}, we see that
this statement also applies to \eqref{eq:KL} whenever $(1-\gamma)\kappa < 1$. For other values of $\kappa$, some conditions on the sample are 
needed to ensure existence for these penalized Tyler's M-estimators \cite{Ollila-Tyler:2014}.  

Curiously, the above conditions on $\kappa$ for a general weight function are needed to establish that the corresponding penalized M-estimator
does not breakdown.  Specifically, suppose now that $\psi(s) = su(s)$ is bounded above by $\kappa$, and so $\psi(s) \le \psi(\infty) = \kappa$.
Let $\widehat{\Sigma}_1$ and $\widehat{\Sigma}_2$ correspond to the unique solution to \eqref{eq:OT} and \eqref{eq:KL} respectively.
In the theorem below, the inequalities refer to the partial ordering of symmetric matrices, i.e.\ $\Sigma \ge \Sigma_o$ if and only if
$\Sigma - \Sigma_o \ge 0$.

\begin{theorem} \label{Thrm:bp}
Under the stated assumptions on $u(s)$, regardless of the data  $X = \{x_1,\cdots, x_n\}$,
\[ \eta I_q \le \widehat{\Sigma}_1 \le \frac{\eta}{1-\kappa} I_q, \] 
provided $\kappa < 1$, and 
\[ \gamma I_q \le \widehat{\Sigma}_2 \le \frac{\gamma}{1-(1-\gamma)\kappa} I_q, \]
provided  $\kappa < 1/(1-\gamma)$.
Hence, under these conditions, the estimators $\widehat{\Sigma}_1$ and $\widehat{\Sigma}_2$ cannot break down, i.e.\
 $\epsilon^*(X;\widehat{\Sigma}_1) = \epsilon^*(X;\widehat{\Sigma}_2) = 1$ when $\eta > 0$ and $0 < \gamma \le 1$ respectively.
 \end{theorem}
 
\noindent
\emph{Proof:}
It is immediate from \eqref{eq:OT} that $\widehat{\Sigma}_1 \ge \eta I_q$. To obtain an upper bound, express \eqref{eq:OT} as
\begin{align*}
\Sigma &= \n\psi\left(x_i^{\tran}\Sigma^{-1}x_i\right)\frac{x_ix_i^{\tran}}{x_i^{\tran}\Sigma^{-1}x_i}+\eta I_q \\
&\le \frac{\kappa \lambda_1}{n} \sum_{i=1}^n \frac{x_ix_i^{\tran}}{x_i^{\tran}x_i}+\eta I_q \le (\kappa \lambda_1+\eta) I_q,
\end{align*}
where $\lambda_1$ denotes the largest eigenvalue of $\Sigma$.  This then implies $\lambda_1 \le \kappa \lambda_1+\eta$ or
$\lambda_1 \le \eta/(1-\kappa)$. Thus,  $\widehat{\Sigma} \le \frac{\eta}{1-\kappa} I_q$. The results for $\widehat{\Sigma}_2$ follow
by relating equations \eqref{eq:OT} and \eqref{eq:KL}.
\hfill $\square$

\section{Hybrid M-estimators} \label{Sec:hybrid}
By construction, the penalized M-estimators obtained from \eqref{eq:OT} and \eqref{eq:KL} are bounded away from singularity for any data set.
We show in this section that even if we correct for this shift, then the resulting estimator still retains a high breakdown point. 

We first focus on \eqref{eq:KL}.  Define $\widehat{V}_2 = (\widehat{\Sigma}_2 - \gamma I_q)/(1-\gamma)$, where $\widehat{\Sigma}_2$ 
denotes the solution to \eqref{eq:KL}, and hence  $\widehat{V}_2$ is the unique solution to 
the M-estimating equation:
\begin{equation} \label{eq:hyb}
V=\n u\left(x_i^{\tran} \{(1-\gamma) V + \gamma  I_q\}^{-1} x_i\right)x_ix_i^{\tran}
\end{equation}
When $\gamma = 0$, \eqref{eq:hyb} corresponds to the usual M-estimating equations for a scatter matrix. On the other hand
when $\gamma = 1$, we obtain
\[ \widehat{V}_2 =\n u(x_i^{\tran}x_i)x_ix_i^{\tran}, \]
which corresponds to a weighted covariance matrix with the data points downweighted based on their Euclidean distance from the center.
These weighted covariance matrices have recently been studied in \cite{Rousseeuw:2019}, wherein they are referred to as general spatial sign 
covariance matrices.  Note that for $u(s)=1/s$ one obtains the usual SSCM. 
  
Consider now the solution to \eqref{eq:OT},  denoted by $\widehat{\Sigma}_1$, and define 
$\widehat{V}_1 = \widehat{\Sigma}_1 - \eta I_q$, which is the unique solution to the M-estimating equation:
\begin{equation} \label{eq:hyb2}
V=\n u\left(x_i^{\tran} (V + \eta  I_q)^{-1} x_i\right)x_ix_i^{\tran}.
\end{equation}
When $\eta = 0$, this corresponds to  the usual M-estimating equations for a scatter matrix. The relationship between \eqref{eq:hyb2} and the
general SSCM is more involved.  Suppose we consider the family of weight functions $u(s) = w(\eta s)$ for a fixed function $w(\cdot)$. Then
as $\eta \to \infty$, we have
$\widehat{V}_1 \to \n w(x_i^{\tran}x_i)x_ix_i^{\tran}$.

The estimators $\widehat{V}_1$ and $\widehat{V}_2$ can thus be viewed as hybrids between a usual M-estimator of scatter and a general 
spatial sign covariance matrix. Rather than choose either extreme as one's estimator,  \eqref{eq:hyb} and \eqref{eq:hyb2} provide classes of scatter 
estimators with tuning parameters $0 \le \gamma \le 1$ and $\eta \ge 0$ respectively. One can then use data driven methods, such as cross-validation, for 
selecting the tuning constant. Moreover, the following theorem shows these hybrid M-estimators have the same breakdown properties as the SSCM.

\begin{theorem} \label{Thrm:bp2}
Suppose $u(s) > 0$ is non-increasing and $\psi(s) = su(s)$ is non-decreasing and bounded by $\kappa = \psi(\infty)$. 
Let $\widehat{V}_1 = \widehat{\Sigma}_1 - \eta  I_q$ and $\widehat{V}_2 = (\widehat{\Sigma}_2 - \gamma I_q)/(1-\gamma)$,
with $\widehat{\Sigma}_1$ and  $\widehat{\Sigma}_2$  being defined as in Theorem \ref{Thrm:bp}.  Furthermore,
suppose the data  $X=\{x_1, \cdots, x_n\}$ spans $\R^q$, and is centered by a known or fixed center. Then, 
\begin{itemize}
\item[i)] for  $\eta > 0$, \ $\epsilon^*(X;\widehat{V}_1) = 1$ \ provided \ $\kappa < 1$. 
\item[ii)] for  $\max\{0, (\kappa - 1)/\kappa\} < \gamma \le 1$,  \ $\epsilon^*(X;\widehat{V}_2) = 1$.
\end{itemize}
\end{theorem}
\emph{Proof:} Consider the contaminated data set \mbox{$Z = X \cup Y$}, where $Y = \{y_1, \ldots, y_m\}$. For this data set, denote 
the solution to \eqref{eq:hyb2} by $\Vhh$, and let $\lambda_q$ be the smallest eigenvalue of $\Vhh$. Since 
\[z^{\tran}\{\Vhh+\eta I_q\}^{-1}z \le z^{\tran}z/(\lambda_q+ \eta),\] 
it follows from the non-increasing property of the weight function $u(s)$ that 
\[ u(z^{\tran}\{\Vhh+\eta I_q\}^{-1}z) \ge u\left(z^{\tran}z/(\lambda_q+ \eta)\right).\]
Thus, from \eqref{eq:hyb2} we obtain
\begin{align*}
  \Vhh  & \ge \frac{1}{n+m} \sum_{i=1}^{n+m} u\left(\frac{z_i^{\tran}z_i}{\lambda_q + \eta}\right)z_iz_i^{\tran}\\ 
    & = \frac{\lambda_q+ \eta}{n+m}\sum_{i=1}^{n+m}\psi\left(\frac{z_i^{\tran}z_i}{\lambda_q+\eta}\right)\frac{z_iz_i^{\tran}}{z_i^{\tran}z_i} \\
        & \ge \frac{\lambda_q+ \eta}{n+m}\sum_{i=1}^{n}\psi\left(\frac{x_i^{\tran}x_i}{\lambda_q+\eta}\right)\frac{x_ix_i^{\tran}}{x_i^{\tran}x_i} \end{align*}
Therefore, $\Vhh$ is bounded below over all possible contaminations $Y$ as long as the good data $X$ spans $\R^q$. By Theorem \ref{Thrm:bp}
it follows that $\Vhh$ is also bounded above over all $Y$. Hence it cannot break down. The proof for $\widehat{V}_2$ is analogous. 
\hfill $\square$

\begin{remark} \label{Rem:center}
If the data is centered by a location statistic $\widehat{\mu}$ having breakdown point $\epsilon^*(X;\widehat{\mu})$, then it can be shown that 
$\epsilon^*(X;\widehat{V}_1) = \epsilon^*(X;\widehat{V}_2) = \epsilon^*(X;\widehat{\mu})$.
\end{remark}

\begin{remark} \label{Rem:comp}
The hybrid M-estimators of scatter can be computed via a simple re-weighing algorithm. Using the results for the fixed-point 
algorithm for finding the solution to \eqref{eq:OT} given in \cite{Ollila-Tyler:2014}, it readily follows that given any initial starting value $V_0 > 0$,
the following algorithms 
 \[ V_{k+1}=\n u\left(x_i^{\tran} (V_k + \eta  I_q)^{-1} x_i\right)x_ix_i^{\tran} \ \mbox{and} \]
  \[ V_{k+1}=\n u\left(x_i^{\tran} ((1-\gamma)V_k + \gamma  I_q)^{-1} x_i\right)x_ix_i^{\tran} \] 
 always converge to  the unique solutions to \eqref{eq:hyb2} and \eqref{eq:hyb} respectively.  
 \end{remark}

\section{Concluding Remarks} \label{Sec:conclude}
An intuitive motivation for considering the hybrid M-estimators of scatter is to observe that the  SSCM does not take into account the shape of the data
cloud when down-weighing observations, but rather down-weights  based on their Euclidean distances from the center.  On the other hand,  the 
non-penalized M-estimators down-weights observations based on their adaptive Mahalanobis distances. However, in addition to  their relatively low breakdown
points in higher dimension, these M-estimators may be ill-conditioned when the sample size is of the same order as the dimension of the data. 

The hybrid M-estimators of scatter can be seen as a compromise between these approaches to down-weighing observations. They take into account the 
shape of the data cloud, but adjust the Mahalanobis contours towards a more spherical shape. This helps prevent problems which may arise from 
inverting an ill-conditioned scatter matrix.  
 
Simulations are ongoing in order to obtain insight into the performance of the hybrid M-estimators, their dependence on the tuning parameter and 
on the choice of the weight function $u(s)$.  The tuning parameter itself may be chosen via data driven methods such as cross validation, or via 
theoretical considerations. Some results for tuning the penalized M-estimator \eqref{eq:OT} for efficiency can be found in \cite{Ollila-Tyler:2014} 
for the case $u(s) = \kappa/s$, with these results applicable to the corresponding hybrid M-estimator. How tuning affects the robustness of
the estimator, though, needs further study.

\section*{Acknowledgements}
Research for both authors was supported in part by National Science Foundation Grants DMS-1407751
and DMS-1812198. Mengxi Yi’s research was also supported in part through a post-doctoral fellowship at
the Technical University of Vienna and in part by the Scientific Research Starting Foundation of UIBE.

\end{document}